\documentclass{article}

\usepackage{arxiv}

\usepackage[utf8]{inputenc} % allow utf-8 input
\usepackage[T1]{fontenc}    % use 8-bit T1 fonts
\usepackage{hyperref}       % hyperlinks
\usepackage{url}            % simple URL typesetting
\usepackage{booktabs}       % professional-quality tables
\usepackage{amsfonts}       % blackboard math symbols
\usepackage{nicefrac}       % compact symbols for 1/2, etc.
\usepackage{microtype}      % microtypography
\usepackage{lipsum}
\usepackage{graphicx}
\graphicspath{ {./images/} }

\usepackage{xcolor}
\usepackage{amssymb}
\usepackage{soul}
\usepackage{subfig}

\usepackage[utf8]{inputenc} % allow utf-8 input
\usepackage[T1]{fontenc}    % use 8-bit T1 fonts
\usepackage{hyperref}       % hyperlinks
\usepackage{url}            % simple URL typesetting
\usepackage{booktabs}       % professional-quality tables
\usepackage{amsfonts}       % blackboard math symbols
\usepackage{nicefrac}       % compact symbols for 1/2, etc.
\usepackage{microtype}      % microtypography
\usepackage{lineno}
\usepackage{float}
\usepackage{hyperref}
\usepackage{url}

\usepackage{graphicx}
\usepackage{xcolor}

\usepackage{url}
\usepackage{ntheorem}

\usepackage[parfill]{parskip}
\usepackage{multirow}
\usepackage{booktabs} 
\usepackage{commath}
\usepackage{amssymb}
\usepackage[utf8]{inputenc}
\usepackage[english]{babel}
\usepackage{mathtools}

\usepackage{soul}
\usepackage{algorithm}
\usepackage{algorithmic}

\usepackage{tipa}
\usepackage{amssymb}

\usepackage{graphicx}
\usepackage{multirow}
\usepackage{gensymb}
\usepackage{amsmath}
\usepackage{hyperref}
\usepackage{times}
\usepackage[utf8]{inputenc} % allow utf-8 input
\usepackage[T1]{fontenc}    % use 8-bit T1 fonts
\usepackage{hyperref}       % hyperlinks
\usepackage{url}            % simple URL typesetting
\usepackage{booktabs}       % professional-quality tables
\usepackage{amsfonts}       % blackboard math symbols
\usepackage{nicefrac}       % compact symbols for 1/2, etc.
\usepackage{microtype}      % microtypography

\usepackage{xcolor}
\usepackage{amssymb}
\usepackage{soul}
\usepackage{subfig}
\usepackage{tabularx}
\usepackage{scrextend}

\usepackage[sort, compress,numbers]{natbib}
\title{Quantum Mechanics and Machine Learning Synergies: Graph Attention Neural Networks to Predict Chemical Reactivity}

\author{
 Mohammadamin Tavakoli\\
  Department of Computer Science\\
  University of California, Irvine\\
  \texttt{mohamadt@uci.edu} \\
   \And
 Aaron Mood\\
  Department of Chemistry\\
  University of California, Irvine\\
  \texttt{amood@uci.edu} \\
  \And
  David Van Vranken\\
  Department of Chemistry\\
  University of California, Irvine\\
  \texttt{David.VV@uci.edu} \\
  \And
  Pierre Baldi \footnote{Corresponding author}\\
  Department of Computer Science\\
  University of California, Irvine\\
  \texttt{pfbaldi@uci.edu} \\
}

\begin{document}
\maketitle

\begin{abstract}
There is a lack of scalable quantitative measures of reactivity for functional groups in organic chemistry. Measuring reactivity experimentally is costly and time-consuming and does not scale to the astronomical size of chemical space. In previous quantum chemistry studies, we have introduced Methyl Cation Affinities (MCA*) and Methyl Anion Affinities (MAA*), using a solvation model, as quantitative measures of reactivity for organic functional groups over the broadest range.
Although MCA* and MAA* offer good estimates of reactivity parameters, their calculation through Density Functional Theory (DFT) simulations is time-consuming. To circumvent this problem, we first use DFT to calculate MCA* and MAA* for more than 2,400 organic molecules thereby establishing a large dataset of chemical reactivity scores. We then design deep learning methods to predict the reactivity of molecular structures and train them using this curated dataset in combination with different representations of molecular structures.
Using ten-fold cross-validation, we show that graph attention neural networks applied to informative input fingerprints produce the most accurate estimates of reactivity, achieving over 91\% test accuracy for predicting the MCA* $\pm 3.0$ or  MAA* $\pm 3.0$,
over 50 orders of magnitude. Finally, we demonstrate the application of these reactivity scores
to two tasks: (1) chemical reaction prediction; (2) combinatorial generation of reaction mechanisms.
The curated dataset of MCA* and MAA* scores is available through the ChemDB chemoinformatics web portal at \url{www.cdb.ics.uci.edu}.
\end{abstract}

\section{Introduction}

In general terms, the chemical reactivity of an atom in a molecule is its propensity towards being an electron donor or acceptor in a chemical reaction. 
Being able to assign reactivity scores to atoms and molecules can be useful
to better understand chemical reactions and their mechanisms in different areas such as chemical synthesis, atmospheric chemistry, drug design, and materials sciences.
Reaction rates have been measured experimentally for a long time, but this is typically a time-consuming process, which becomes increasingly costly as one tries to explore the most challenging reactions. Moreover, true solution-phase reaction rates are bounded by the rates of molecular diffusion, complicating the quantifying the extremes of reactivity.
Herbert Mayr and his colleagues have pioneered the empirical study of chemical reactivity
by laboriously measuring the reactivity of the main organic functional groups and deriving corresponding scales of reactivity \citep{mayr2008general,mayr2015quantitative}. However, due to the experimental limitations, their scale covers only a limited range of electrophiles and nucleophiles.
An alternative approach to derive reactivity scores is to use quantum mechanical (QM) simulations.

\begin{figure}[!]
  \centering
  \includegraphics[width=\linewidth]{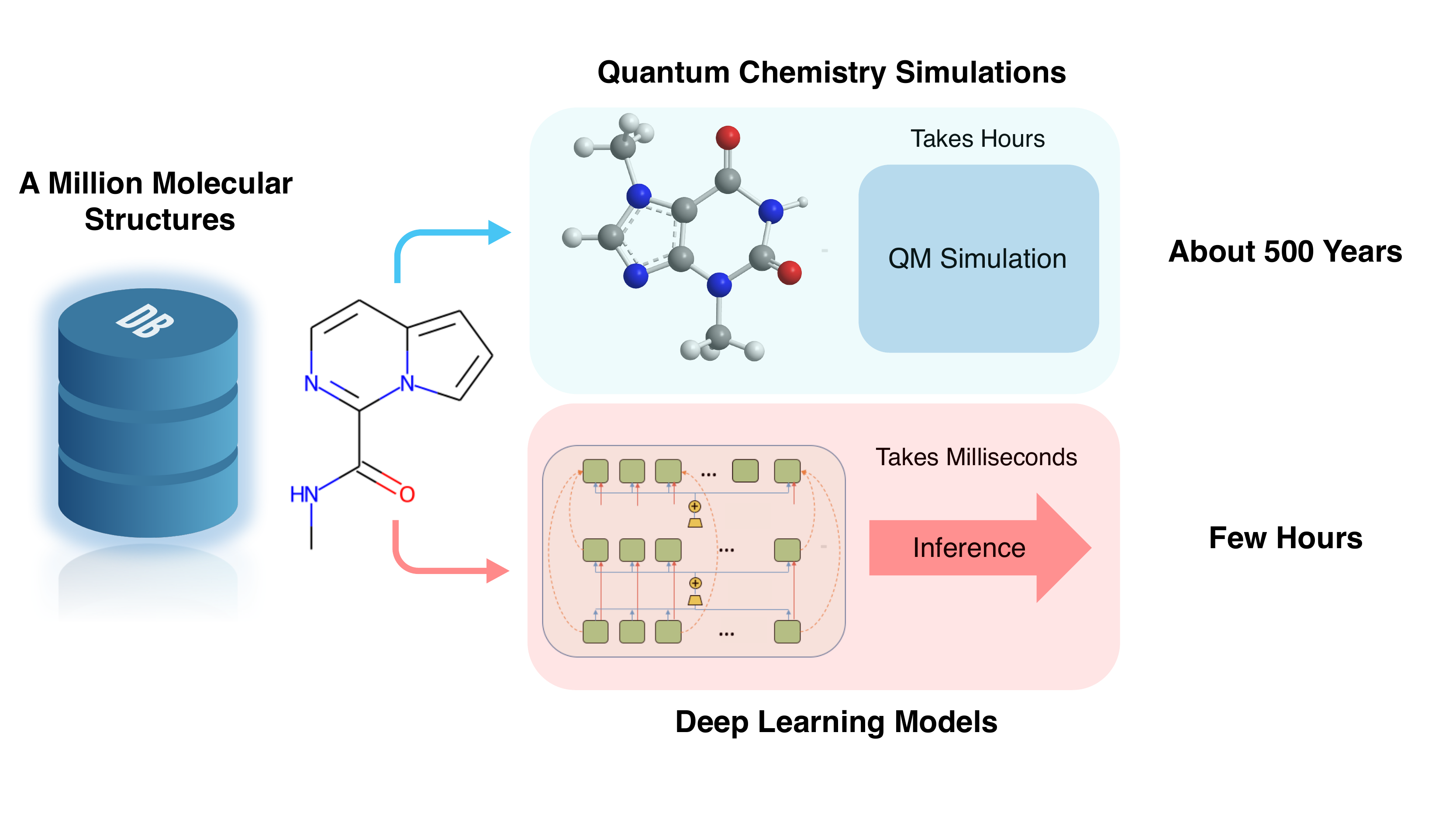}
  \caption{Typical speed differences in computing a given molecular property for a database of 1M molecules using DFT versus machine learning. Assuming QM simulations take approximately five hours per molecular structure, the total processing time for 1M structures is $5 \times 3600\text{(s)} \times 10^6 \approx 500 \text{ years}$. This processing time can be reduced to a few hours, after training a suitable neural architecture, assuming that inference time in a trained deep learning model is in the range of $5$ milliseconds: $5 \times 10^{-3}\text{(s)} \times 10^6 \approx \text{few hours}$.}
  \label{fig:time}
\end{figure}

Recently, we used QM with Density Functional Theory (DFT) simulations to investigate this problem for electron donor and electron acceptor functional groups \citep{mood2020methyl, mca2021}. Specifically, we applied DFT to over 100 diverse molecular structures and showed that in general methyl ion affinities, using a solvation model, are highly correlated to the Mayr reactivity scale. However, while using QM is faster than running laboratory experiments and can potentially cover a larger range of electrophiles and nucleophiles, the underlying DFT calculations still take up several hours for a molecule with only twenty atoms.
Therefore, here we develop a more efficient approach leveraging the synergies between QM and machine learning \citep{sadowski2016synergies}, where we first use DFT to produce a substantial training set of chemical reactivity scores and then develop and train machine learning methods to interpolate and extrapolate chemical reactivity scores in real-time. These machine learning methods, in particular deep learning methods, have
already been successfully applied to a variety of chemoinformatics problems, including the prediction of molecular properties
\citep{ralaivola05b,lusci2013deep,lusci2015accurate}
and reactions \citep{kayala2011learning,kayala2012reactionpredictor,coley2017prediction, fooshee2018deep, de2018molgan,you2018graph, zhou2019optimization, coley2019graph, tavakoli2020continuous}. Training these models may take some time but, once trained, they are orders of magnitude faster than QM calculations and can generalize, making them a suitable complement to time-consuming DFT simulations (Figure \ref{fig:time}). 

%%%%%%%%%%%%%%%%%%%%%%%%%%%%%%%%%%%%%%%%%%%%%%%%%%%%%%%%%
 %%%%%%%%%%%%%%%%%%%%%%%%%%%%%%%%%%%%%%%%%%%%%%%%%%%%%%%%%
In what follows, we first explain the concept of reactivity and how it correlates with methyl ion affinities. Then we describe the curated dataset we produce using DFT calculations. We then develop different machine learning methods tuned to different molecular representations and interpretations of molecular reactivity and proceed to train, test, and compare them using the curated data set. Finally, we discuss the potential applications of this approach for estimating the relative reactivity of atoms for the tasks of chemical reaction prediction and combinatorial generation of organic mechanisms.

\vspace{2.5mm}

\section{Reactivity Scales}
\vspace{2.5mm}
Until the pioneering work of Herbert Mayr and his team, the very idea
that nucleophilicity or electrophilicity might be quantified on independent scales had eluded chemists.
Through a massive experimental and theoretical undertaking, Mayr and his collaborators have made comprehensive and systematic measurements of reaction rates of various electrophiles and nucleophiles in the laboratory. Furthermore, they have shown that the solution-phase nucleophilicity $N$ and electrophilicity $E$ can be independently quantified using a logarithmic scale that correlates with the free energy of activation, allowing useful predictions of reaction rate constants, below diffusion control, using the now-famous equation: $\log_{10} k_{25\degree} = (E + N)s_N$, where $s_N$ is a nucleophile-dependent parameter typically near unity and $k$ is the rate constant. The success of this equation centers around a focus on reactions that form bonds to carbon atoms; given the importance of solution-phase organic chemistry in biochemistry and medicine, this is a reasonable restriction 
 \citep{mayr2008general,mayr2015quantitative}.

Methyl cation affinity (MCA) is defined as the energy difference resulting from combining a methyl cation with a nucleophile; similarly, methyl anion affinity (MAA) is defined as the energy difference resulting from combining a methyl anion with an electrophile (Figure \ref{fig:mca_maa}). Mayr, Ofial, Zipse, and others have shown that MCA and MAA correlate with the experimentally measured nucleophilicities and electrophilicities on a relatively small range of reactivity \citep{schindele2002relationships,seeliger2008reactions,troshin2011electrophilicities,allgauer2017quantification,bottger2000electrophilic}($\sim 30$ orders of magnitude \citep{pellerite1983intrinsic,wei2008methyl}).

Recently, Van Vranken and his collaborators have shown that calculated methyl cation affinities and methyl anion affinities, with the inclusion of a solvation model, (MCA*) and (MAA*) where * denotes the solvation model, are highly correlated with measured nucleophilicity $N*s_N$ and electrophilicity $E$ over a broad range of molecules containing first and second-row atoms \citep{mood2020methyl, mca2021}. They used this correlation to expand the lower and upper ends of the nucleophilicity and electrophilicity scale produced by Mayr.
Their work introduced MCA* and MAA* as new metrics to estimate reactivity parameters without carrying laboratory experiments. Additionally, they showed that MCA* and MAA* provide useful reactivity scores in a much broader range of chemical reactivity than previous work (up to $\sim 180$ orders of magnitude). 
Since calculating MCA* or MAA* only requires quantum chemistry simulations, this approach is relatively fast compared to experimental approaches. However, it still requires extensive computational resources and time-consuming processes and cannot be used in high-throughput mode, hence the need to develop faster approaches. 

The resulting chemical reactivity scores (MCA* and MAA*), formulated as the difference between the energy of reactants and products of the reactions shown in Figure \ref{fig:mca_maa}, can be interpreted in slightly different ways, depending on the entity to which the score is attributed to. Van Vranken et al. interpreted these quantities as the electrophilicity and nucleophilicity of the reacting functional groups \citep{mood2020methyl, mca2021}. However, in cases where only one functional group is going to react with methyl ions, the corresponding values can be attributed to the reactivity of the entire molecule. Another possibility is to attribute the MCA* and MAA* scores to individual atoms. As shown in Figure \ref{fig:mca_maa}, during the reaction with methyl ions, the ion is attaching to one particular atom of the electrophilic or nucleophilic molecule. This atom is labeled as the most reactive atom within the electrophilic or the nucleophilic group, so one can assign the MCA* or MAA* as the relative reactivity of that specific atom. Although all these interpretations are valid and will be further examined, in most of what follows we will use the third attribution at the level of atoms. Throughout the rest of this paper, MCA* and MAA* are referred to as the nucleophilicity (electron donor) and electrophilicity scores (electron acceptor).

\vspace{2.5mm}
\section{Curated Dataset of Chemical Reactivities}
\begin{figure}[h]
    \centering
    \subfloat[][\centering]{\includegraphics[ width=6.5 cm]{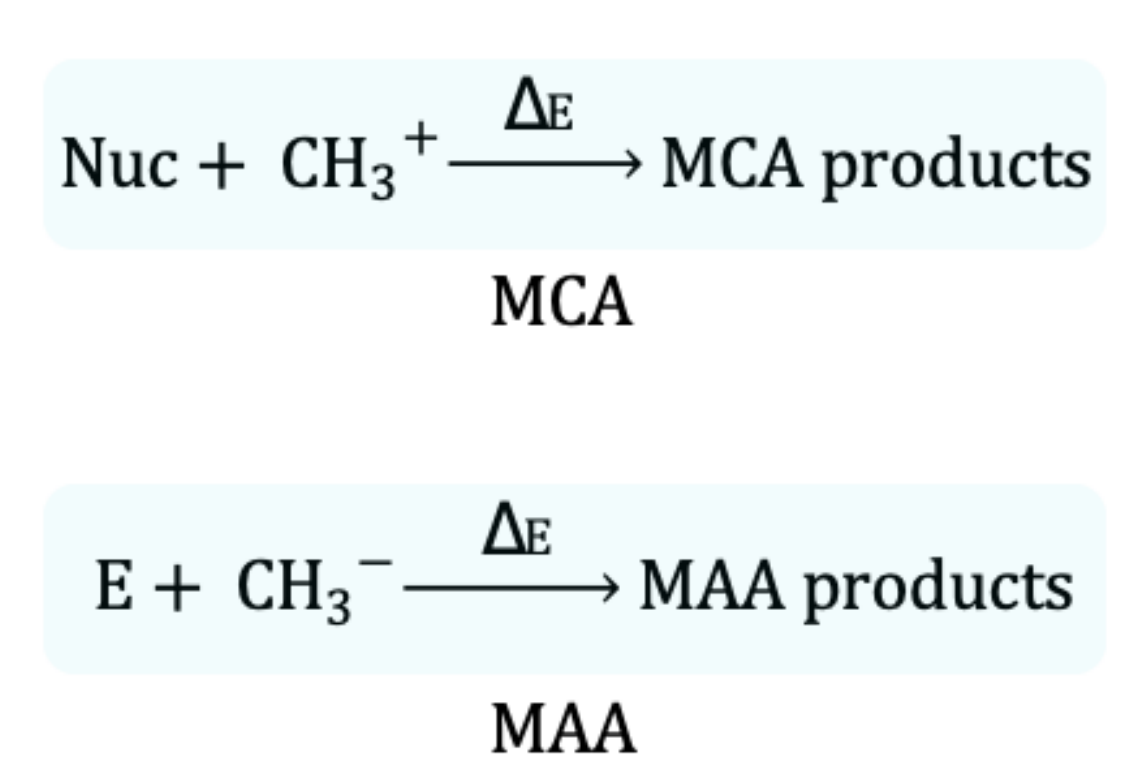}\label{fig:mca_maa}}%
    \qquad
    \subfloat[][\centering ]{\includegraphics[width=8.5cm]{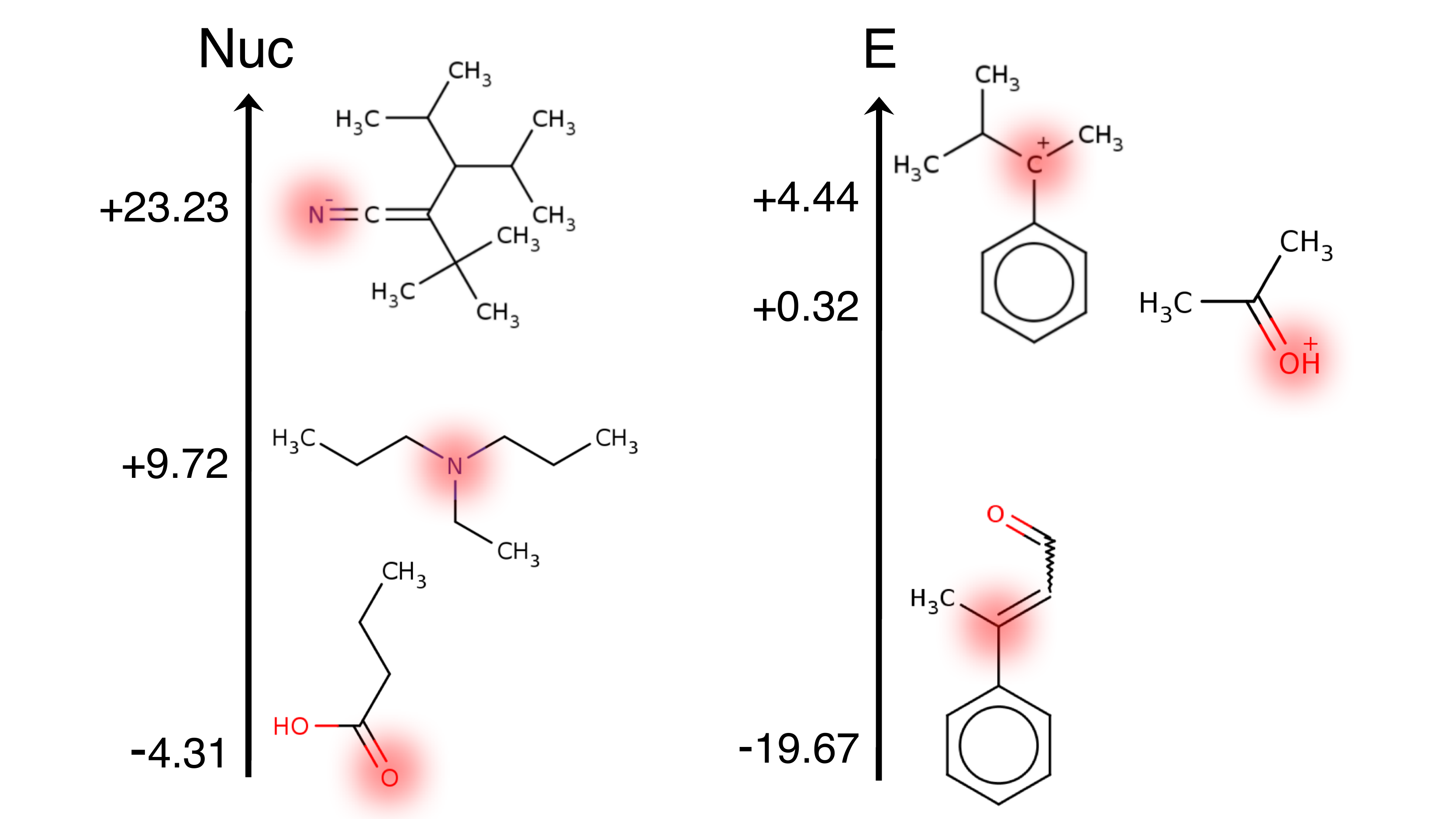}\label{fig:data_sample}}%
    \caption{(a): Schematic definition of {MCA} and {MAA}. Nuc = Nucleophile. E =  Electrophile. 
  MCA = The negative of the energy difference resulting from combining a methyl cation with a nucleophile.
  MAA = The negative of the energy difference resulting from combining a methyl anion with an electrophile.
  (b): A few molecules from the curated dataset and the corresponding reactivity scores shown on two separate scales for nucleophilicity and electrophilicity.
  }%
    \label{fig:fig2}%
\end{figure}

\vspace{2.5mm}
Following the method for calculating chemical reactivity in \citet{mood2020methyl, mca2021}, we compute the methyl cation affinities (MCA*) of 1232 nucleophiles and the methyl anion affinities (MAA*) of 1189 electrophiles. 
These molecules contain simple carbon skeleton structural variations to improve the
generalizability of the trained models such as the ones shown in Figure \ref{fig:data_sample}. The nucleophilic functional groups include: amines, ethers, amide anions, alkyl carbanions, aldehydes, ketones, esters, carboxylic acids, amides, enolates, nitronate anions, diazo compounds, cyanoalkyl anions, imines, nitriles, isonitriles, and bis(cyano)alkyl anions. The electrophilic functional groups include: iminium ions, imines, oxonium ions, aldehydes, ketones, esters, amides, benzyl cations, allyl cations, alkyl cations, carbonyl Michael acceptors, nitrile Michael acceptors, and nitro Michael acceptors. Methyl ion affinities (MCA* and MAA*) were calculated using TURBOMOLE V7.3 at the PBE0(disp)/DEF2-TZVP COSMO($\infty$) level of theory \citep{li2018kinetics,grimme2011effect}.
It must be noted that the MCA* and MAA* scores
are correct within a range of three orders of magnitude \citep{mood2020methyl, mca2021}, e.g., the reported number for C[CH:1]=[O+]C is 3.29 which means the actual reactivity score is within the range of 0.29 to 6.29.

\iffalse
Using the synthetic dataset of RRS we design and train deep learning models to estimate the chemical reactivity scores.
In the following section, we explain the details about the deep learning models and the corresponding performance in predicting RRS for unseen chemical structures.
These numbers are 
The dataset of relative reactivity scores is available to download through ICS ChemDB web portal at www.cdb.ics.uci.edu
\fi

%%%%%%%%%%%%%%%%%%%%%%%%%%%%%%%%%%%%%%%%%%%%%%%%%%%%%%%%%

%%%%%%%%%%%%%%%%%%%%%%%%%%%%%%%%%%%%%%%%%%%%%%%%%%%%%%%%%
\vspace{2.5mm}
\section{Deep Learning to Predict Chemical Reactivity}
\vspace{2.5mm}
Deep learning \citep{baldi2021deep} methods have many applications in chemoinformatics,
from molecular property prediction 
\citep{lusci2013deep, lusci2015accurate, feinberg2018potentialnet,tavakoli2020continuous}
and optimization \citep{zhou2019optimization, olivecrona2017molecular, de2018molgan, fmodel},
to reaction prediction 
\citep{kayala2011learning,kayala2012reactionpredictor,coley2017prediction, fooshee2018deep, de2018molgan,you2018graph, zhou2019optimization, coley2019graph, schwaller2019molecular},
to the acceleration of QM calculations \citep{PhysRevLett.108.253002,hermann2020deep}.
Although it takes time to train deep learning methods, at inference time they are fast and tend to generalize well. Deep learning methods in chemoinformatics must be developed in tandem with the underlying chemical representations: while feedforward neural network can be applied to vectorial or tensorial representations of fixed size, recursive or graph neural network must be used in the case of variable size, structured, representations \citep{urban2018inner}.
Next, we describe the representations and architectures we use for the problem of predicting electrophilicity and nucleophilicity.

\vspace{2.5mm}
\subsection{Representation of Molecular Structures}
\vspace{2.5mm}
As previously mentioned, there are several ways of attributing the reactivity score introduced in \citet{mood2020methyl, mca2021}. This score is defined as the difference between the free energy of the reactants and products in the reactions shown in Figure \ref{fig:mca_maa}. It is also calculated for the most reactive atom of the most reactive functional group in each molecule. Therefore, one can interpret this as: (1) the reactivity of the atom which is bonding to the methyl ion during the reaction (atomic reactivity); (2) the reactivity of the functional group which contains the atom that is bonding to the methyl ion (group reactivity);
or (3) the reactivity of the molecule which is reacting with the methyl ion (molecular reactivity). Using our know-how and some preliminary exploration to avoid looking at all possible combinations of attribution and representations, we converged on the following list of possible representations:
(1) informative fingerprint vector representation to learn the atomic reactivity;
(2) extended connectivity fingerprint (ECFP) to learn the molecular reactivity;
(3) SMILES string text representation to learn the molecular reactivity;  and (4) graph representations to learn atomic, functional group, and molecular reactivity. Next, we develop deep learning models congruous with each representation.

\vspace{2.5mm}
\subsection{Deep Learning Models}
\vspace{2.5mm}
\subsubsection{Informative Fingerprint Vector Representation and Model}
To perform an atom level prediction, each atom has to be mapped to a vector.
The first method to find this mapping is to use chemical features associated with each atom in a molecule. These features fall into two categories: 1) graph-topological; and 2) physical-chemical.
Graph-topological features reflect patterns of connectivity and the neighborhood of atoms, as in standard molecular fingerprint representations. The physical-chemical features instead capture properties of the atom itself. Examples of physical-chemical features are the presence and type of filled and unfilled orbitals, electronegativity, and the location of the atom in the periodic table.
In this work, we associate a feature vector of length 52 to each atom. This vector corresponds to the concatenation of 44 graph topological features and 8 physical-chemical features.
We refer to this vector as the informative fingerprint vector representation of the corresponding atom. Then we train two separate neural networks, one for electrophilicity prediction, and one for nucleophilicity prediction.
After a hyperparameter optimization phase carried using Sherpa \citep{hertel2018sherpa}, both resulting networks comprise two hidden layers with 32 and 16 units and one output linear unit for the regression task using the mean squared error. We use dropout \cite{srivastava2014dropout,baldi_dropout_2014} on the hidden layers with a rate of 0.4. 
We also use SPLASH activation functions initialized to ReLU activations \citep{tavakoli2021splash} in the hidden layers.  The results of this experiment for both electrophilicity and nucleophilicity are shown in Table \ref{tab:acc_result}.

\vspace{1.5mm}
\subsubsection{Extended Connectivity Fingerprint Representation and Model}
\vspace{1.5mm}
When the reactivity score is attributed to the molecule (molecular reactivity), we use molecular fingerprints to represent the entire molecule as a binary or integer (count) vector. In this work, we use the well-known extended-connectivity fingerprints (ECFP) \citep{rogers2010extended} to train two separate networks for electrophilicity and nucleophilicity prediction.
The length of the fingerprints must be adjusted as there is a basic trade-off: longer fingerprints capture more information, however, they increase the risk of overfitting. Considering the length of the fingerprints and the radius as hyperparameters, after some experimentation, we converge on a size of 512 with a radius of four. 
The same architecture as the one used in the previous section is employed here for both networks, with 64 and 32 units in the hidden layers, followed by a similar linear output unit. We apply $L2$ regularization (with $\lambda = 0.20)$ to the weights.

\vspace{1.5mm}
\subsubsection{SMILES String Representation and Model}
\vspace{1.5mm}
We also use canonical SMILES strings \citep{weininger1988smiles} to represent the molecules.
In order to apply deep learning methods, the SMILES strings must be converted to a numerical format. The most straightforward and widely-used embedding is the character level embedding \citep{liu2017retrosynthetic, ikebata2017bayesian, schwaller2019molecular}.
However, it is more efficient to use atomic symbols as the embedding units. Not only does this reduce the extra computations required by atoms represented with multiple characters, but it also provides a clearer separation between pairs of atoms versus atoms represented by multiple characters (e.g. Sc can be seen as either a Sulfur atom connected to an aromatic carbon or a Scandium atom).
In this embedding, each atom and special character in a SMILES string is mapped into a high dimensional vector. This can be done through an embedding layer whose weights are learned during the training process. However, because we are using a relatively small training dataset of reactivity scores, we avoid adding extra embedding parameters to the model by using a pre-trained embedding of atoms. We use the trained atom embedding vectors used in \citet{fooshee2018deep} for an atom classification task within a reaction prediction pipeline. In this case, atoms are mapped onto a 10-dimensional vector space. 
Figure \ref{fig:atom_tsne} shows the t-SNE 2D visualization of the embedded atoms preserving some of their physical-chemical properties and separating them from special characters. 

To predict the reactivity as the molecular property (i.e. molecular reactivity), the molecules are represented as SMILES strings. Two neural networks with the same architecture are employed to predict electrophilicity and nucleophilicity. The architecture of the neural networks consists of a pre-trained atom embedding layer followed by a one-dimensional convolution layer with a window size of five. This is followed by a bidirectional LSTM layer with 16 hidden units. Then the output is computed by a linear unit fully connected to the previous layer. Due to the increase in the number of parameters, $L2$ regularization with $\lambda=0.20$ is applied for each weight of the convolution and LSTM layers. The results of this experiment are shown in Table \ref{tab:acc_result}.

\vspace{1.5mm}
\subsubsection{Graph Representation and Model}
\vspace{1.5mm}
Finally, there are deep learning methods that can be applied directly to graph-structured data, such as 
knowledge graphs, social networks, parse trees, and molecules
 \citep{lusci2013deep,duvenaud2015convolutional,urban2018inner, schlichtkrull2018modeling, velivckovic2017graph, kearnes2016molecular}. Within these methods, the graph convolutional network (GCN), or outer recursive neural network approach, is 
 particularly suited for processing molecular graphs through an iterative 
message passing mechanism that aggregates information about each atom's
over increasing larger neighborhoods. 
Here we follow an approach similar to \citet{schlichtkrull2018modeling} to model a molecule as a relational graph structure. Each molecule is first represented as an undirected labeled graph $G = (V, E, R, S)$, where the nodes in $V$ correspond to the atoms and the edges in $E$ to the bonds. The labels for the vertices in $R$ correspond to atom types (e.g. C, O). The labels for the edges in $S$ correspond to edge types (single, double, triple, aromatic). 
In the beginning, each node of the graph is associated with a vector representation 
(initial mapping) carrying information about the node. The edges of the graph can be treated in different ways; they can be mapped to real-valued vectors using learnable embedding weights, or they can be mapped to binary vectors using the one-hot encoding of the bond types. Since we are focusing on organic structures, there are only four types of bonds in the data set, and therefore the one-hot encoding approach is both reasonable and economical. Thus a molecule with $n$ atoms and a feature vector of length $d$ for each atom (initial mapping) can be represented by an $n \times d$ node-feature matrix $H^0$ together with an $n \times n \times d'$ adjacency tensor $A$ (here $d'=4$). 
 Each row in the node-feature matrix is the vector representation of the corresponding atom. In the adjacency tensor, for a pair of vertices $i$ and $j$, the corresponding vector of length four represents the one-hot-encoding of the corresponding bond type. For atoms that are not bonded, the corresponding vector is $(0,0,0,0)$. 
 Using this representation, we can apply a graph convolutional neural network to recursively update the representation of each atom. The recursive propagation of information in the convolutional neural network is given by:
\begin{align}
h_i^{l+1} =  h_i^{l} + \sigma(\sum_{j \in N(i)} W^l h_j^l )
\end{align}
Here $h_i^l$ is the vector representation associated with node $i$ at level $l$. $W^l$ denotes the shared weights of the convolution applied from level $l$ to level $l+1$. $N(i)$ denotes the neighborhood of vertex $i$, consisting of its immediate neighbors and $\sigma$ is a non-linear transformation.
In order to take different bond types into account, we use the approach in \citet{schlichtkrull2018modeling} using a different set of weights for each bond type, resulting in the form:
\begin{align}
h_i^{l+1} = h_i^{l} + \sigma(\sum_{k \in S} \sum_{j \in N_k(i)} (W_k^lh_j^l))
\label{relational_GCN}
\end{align}
Where $N_k(i)$ denotes the set of nodes connected to node $i$ through the edge type $k \in S$. 
To train this graph neural network, we optimize a multi-objective loss function.
In addition to minimizing the error between the predicted reactivity and the actual value of MCA* or MAA* for the node of interest, we penalize the network whenever the predicted reactivity of other nodes is greater than the predicted reactivity of the node of interest. This can be formulated as follows:
\begin{align}
\mathcal{L} =  ||f(h^{L}_i) - y||^2 +  \beta [ReLU(max_j((f(h_j^L))-f(h_i^L))]^2
\label{loss}
\end{align}
where $i$ is the index of the atom which reacts with methyl ions, $h^L$ is the atom representation produced by the last convolutional layer of the GCN ($L$), and $f(.)$ is a fully connected linear function from $\mathbb{R}^d$ to $\mathbb{R}$ which outputs the reactivity prediction. The first loss term is the standard least square 
regression loss for predicting the target 
value of the MCA* or MAA* masked to the node of interest $i$. The second term looks at all the other nodes $j$ and the corresponding reactivity prediction $f(h_j^L)$, computes the maximum reactivity difference with respect to node $i$, retains it only in the unwanted case where this difference is positive (through the ReLU function), and applies a square loss with a weighting hyperparameter $\beta$.
In other words, the reactions shown in Figure \ref{fig:mca_maa} have the highest reaction rate among all other plausible reactions with the same set of reactants.  For the initial mapping of each atom ($h_i^0$) we use a one-hot vector encoding of the atom type concatenated with eight physical-chemical features. As can be seen from Equation \ref{relational_GCN}, applying the first level of graph convolution updates the node representation using information from its immediate (one-hop) neighbors. To incorporate information from nodes up to three hops away into an atom's representation, we apply three layers of graph convolution with node representations of length 38, 24, 16, and 8 chosen after some exploratory experimentation.
We also concatenate a count vector of 10 predefined molecular graph connectivity patterns to the output of the top convolutional layer. These predefined patterns have graph lengths greater than three and thus cannot be captured by the three-layer GCN. ReLU activation functions are used within the GCN layers and, after some experimentation, the hyperparameter $\beta$ is set to 0.2. An $L2$ regularization with hyperparameter $\lambda=0.2$ is applied to all the weights to avoid any overfitting. The GCN performance is shown in Table \ref{tab:acc_result}.

There are a number of graph pooling mechanisms, such as those described in \citet{lee2019self, murphy2019relational}, that can be used to predict molecular reactivity and functional group reactivity after the graph convolutions.
Because of our relatively small training dataset, to avoid the risk of overfitting, we try simple pooling mechanisms that require minimal additional learning steps. In addition to the function $f$ described above, we use an element-wise max-pooling, average-pooling, and the pooling mechanism introduced in \citet{duvenaud2015convolutional} which is described in Equation \ref{pooling}, where $G(V)$ is the final representation of the entire graph and $N$ is the number of input nodes to the pooling operation. When predicting molecular reactivity, $N = |V|$; when predicting group reactivity, $N$ is the number of atoms within the functional group which reacts with the methyl ions. As previously seen, $h_i^L$ is a vector of size $n_L$ representing the node $i$ in the top convolutional layer $L$, and $W_p$ is the shared pooling weight matrix of size $(n_L, m)$. Since this pooling is known to work better with relatively longer node representations \citep{duvenaud2015convolutional}, we set $m=n_l$. In our experiments, however, we found that for the prediction of molecular reactivity and group reactivity max-pooling and average-pooling work slightly better than the mechanism introduced in \citet{duvenaud2015convolutional}. 

\begin{align}
\label{pooling}
G(V) = \sum_{i=1}^{N} \frac{\exp(o_i)}{\sum_{1}^{m}\exp(o_i)}, \hspace{5mm} o_i = h_i^L W_p
\end{align}

\subsubsection{Graph Attention Networks}
\vspace{1.5mm}
Another recently introduced type of neural network which can operate on graph data is the Graph Attention Network (GAT) \citep{velivckovic2017graph}. In this approach, node representations are updated by a weighted message passing scheme between neighbors, where the weights are calculated through an attention mechanism as follows:
\begin{align}
h_i^{l+1} = {\vert\vert}_{m=1}^{P}\sigma(\sum_{j\in N_i}^{} \alpha_{(m)}^{(ij)} h_j^l W_{(m)}^l)
\quad {\rm with} \quad 
\alpha_{(m)}^{(ij)} = \frac{\exp(A(W_{(m)}^lh_i \vert\vert W_{(m)}^lh_j))}{\sum_{k\in N_{(i)}}^{}\exp(A(W_{(m)}^lh_i \vert\vert W_{(m)}^lh_k))}
\end{align}
Here $P$ is the number of attention mechanisms (also known as attention heads \citep{velivckovic2017graph}), $m$ ranges from 1 to $P$, $W_{(m)}^l$ denote the shared weights of layer $l$ and attention mechanism $m$, and $A$ represents a shallow neural network with a leaky-ReLU output activation mapping the input vector into a scalar. The symbol $\vert\vert$ represents the concatenation operation and $h_i^{l+1}$, as the updated node representation, is the concatenation of the output of different attention mechanisms.
Through our experiments, we find that using a larger node representation at the last layer of the GAT together with averaging (instead of concatenation) works better. We use three GAT layer representations of size 38, 24, 16, and 12 and concatenate the representation of the top layer with the same count vector of predefined patterns used with the GCN. The GAT is trained using the same loss function as described in Equation \ref{loss}.
We experimented with several variations in terms of the number of attentions heads and bond representations and report only the best results ($P=3$), although the difference observed were minor. The GAT results for the prediction of reactivity are shown in Table \ref{tab:acc_result}.

\iffalse
In the experiments, we compared two GAT architectures: (1) one with three attention heads $P=3$ and to reduce the number of parameters, instead of using different weights for each bond type (i.e. edges), we use a learnable affine transformation of the weights corresponding to single bond as the weights for other bond types. (2) We use $P=1$ with independent weights for each bond type to use approximately the same number of parameters as used in (1). Through our experiments, we observe that the former approach performs slightly better. The results of GAT for the prediction of reactivity are shown in Table \ref{tab:acc_result}.
\fi

\section{Results}

\subsection{Data}
\begin{figure}[h]
    \centering
    \subfloat[][\centering]{\includegraphics[width=7.5cm]{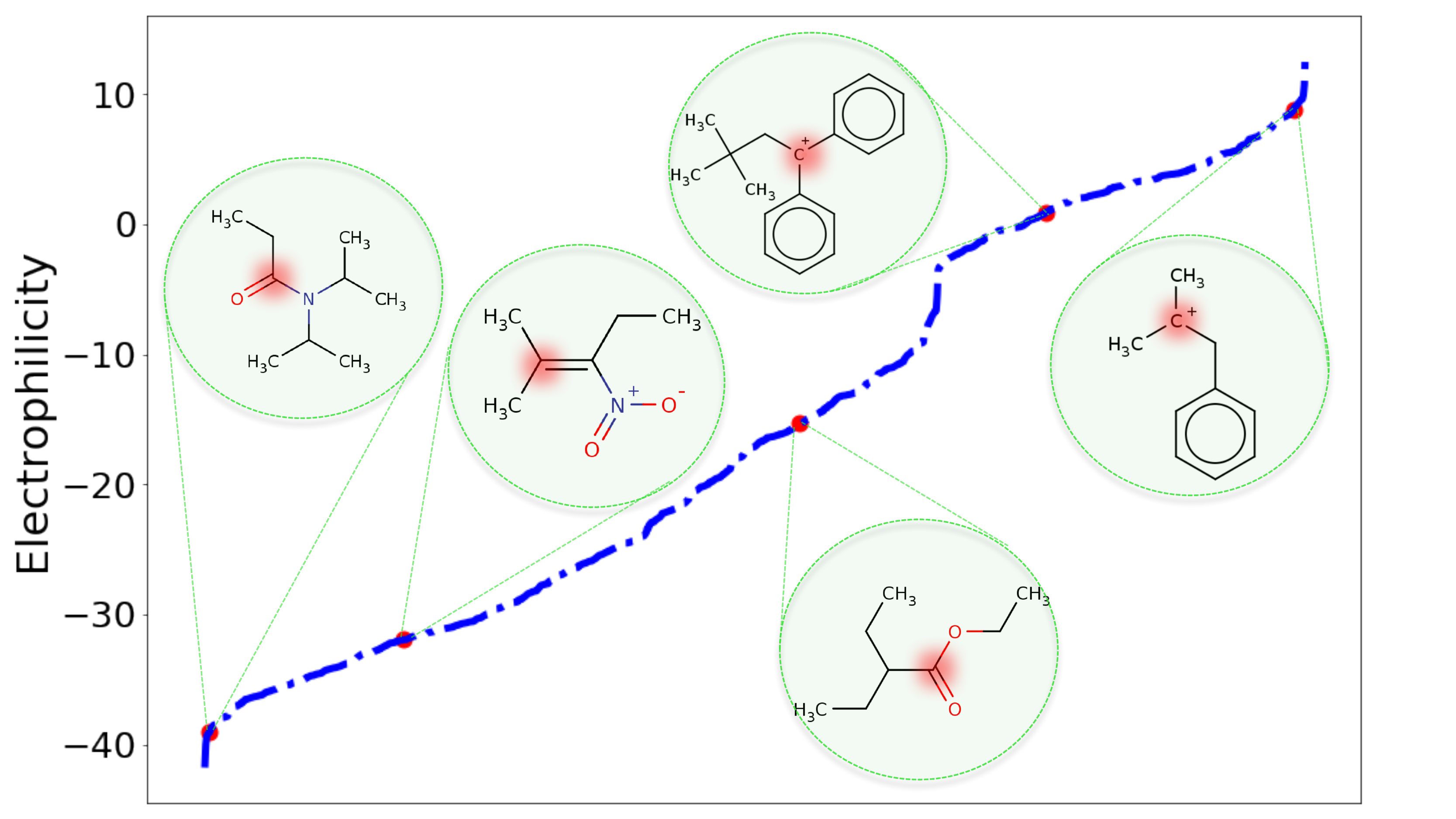}\label{fig:elec_data}}%
    \qquad
    \subfloat[][\centering]{\includegraphics[width=7.5cm]{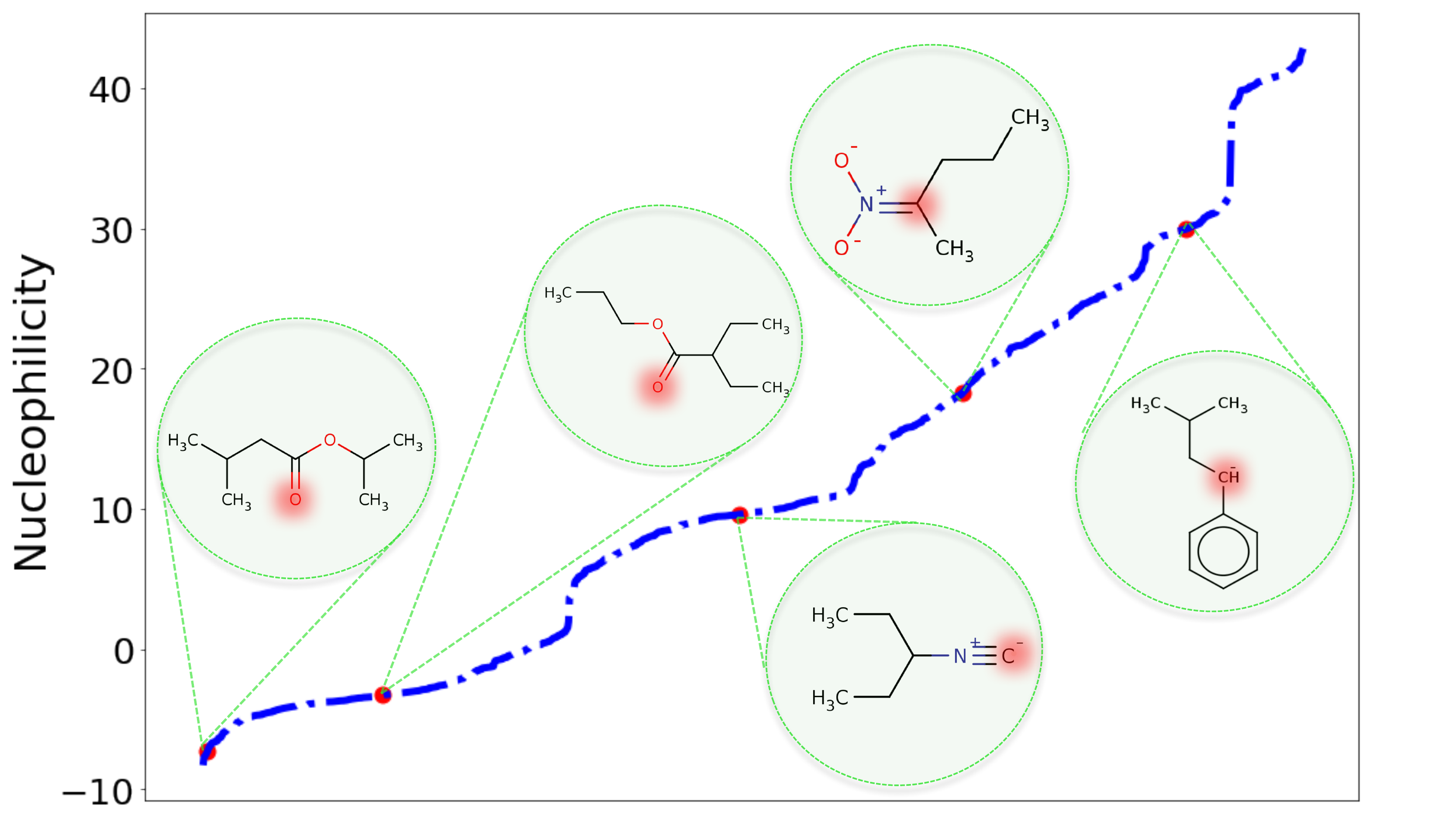}\label{fig:nuc_data}}%
    \caption{The range of electrophilicity and nucleophilicity in the curated dataset, covering $\sim 50$ orders of magnitude in each case, together with five examples of electrophiles and nucleophiles.}%
    \label{fig:train_data}%
\end{figure}

The full curated data set of 2421 molecules and their scores, covering 53 orders of magnitude of chemical reactivity,
is available for download from the UCI ChemDB chemoinformatics web portal at \url{www.cdb.ics.uci.edu}. Figure \ref{fig:train_data} illustrates the range of electrophilicity and nucleophilicity in the curated dataset.

%%%%%%%%%%%%%%%%%%%%%%%%%%%%%%%%%%%%%%%%%%%%%%%%%%%%%%%%%

%%%%%%%%%%%%%%%%%%%%%%%%%%%%%%%%%%%%%%%%%%%%%%%%%%%%%%%%%

\subsection{Comparative Analysis and Predictions}

\begin{figure}[!]
    \centering
    \subfloat[][\centering]{\includegraphics[width=7.5cm]{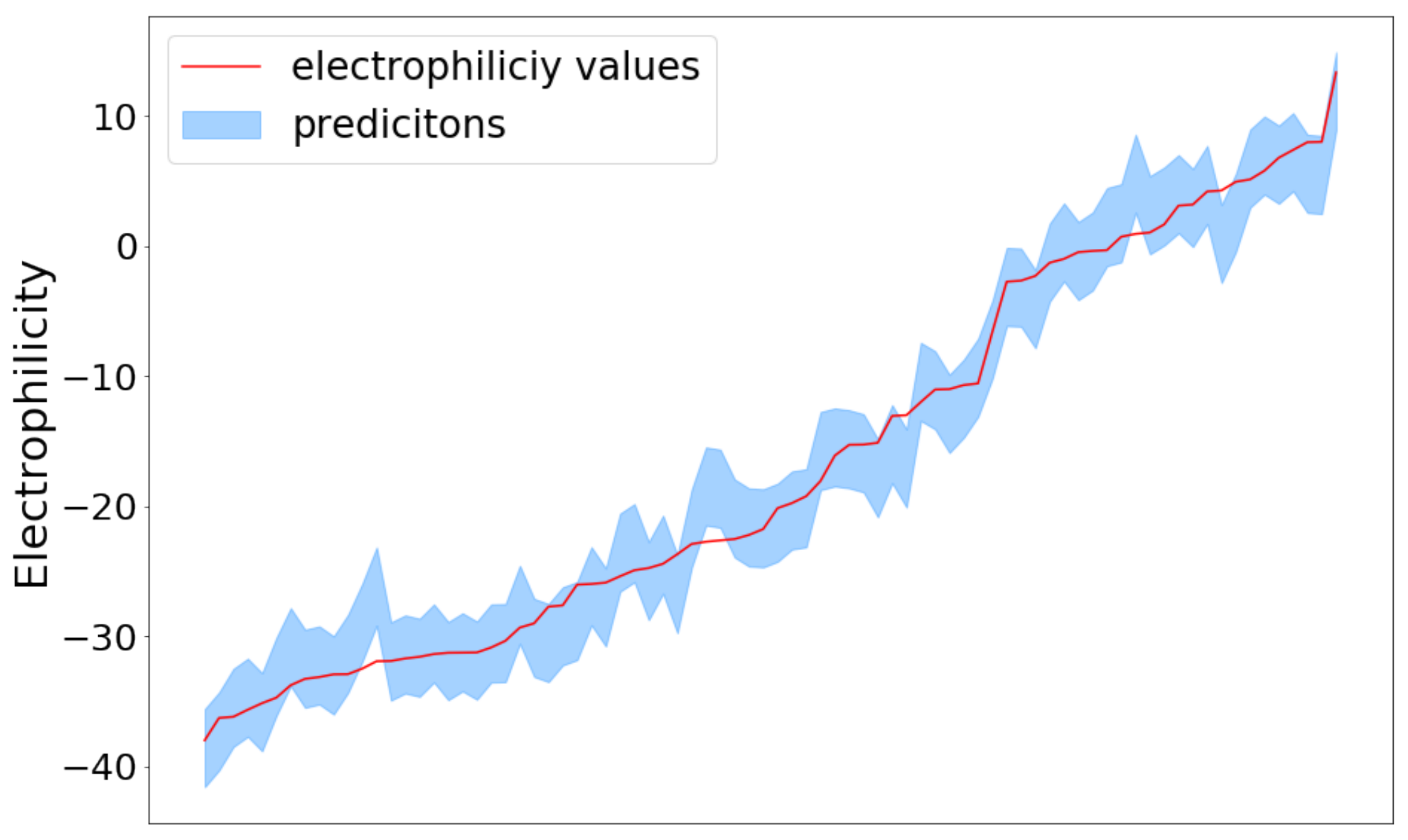}\label{fig:elect_test}}%
    \qquad
    \subfloat[][\centering]{\includegraphics[width=7.5cm]{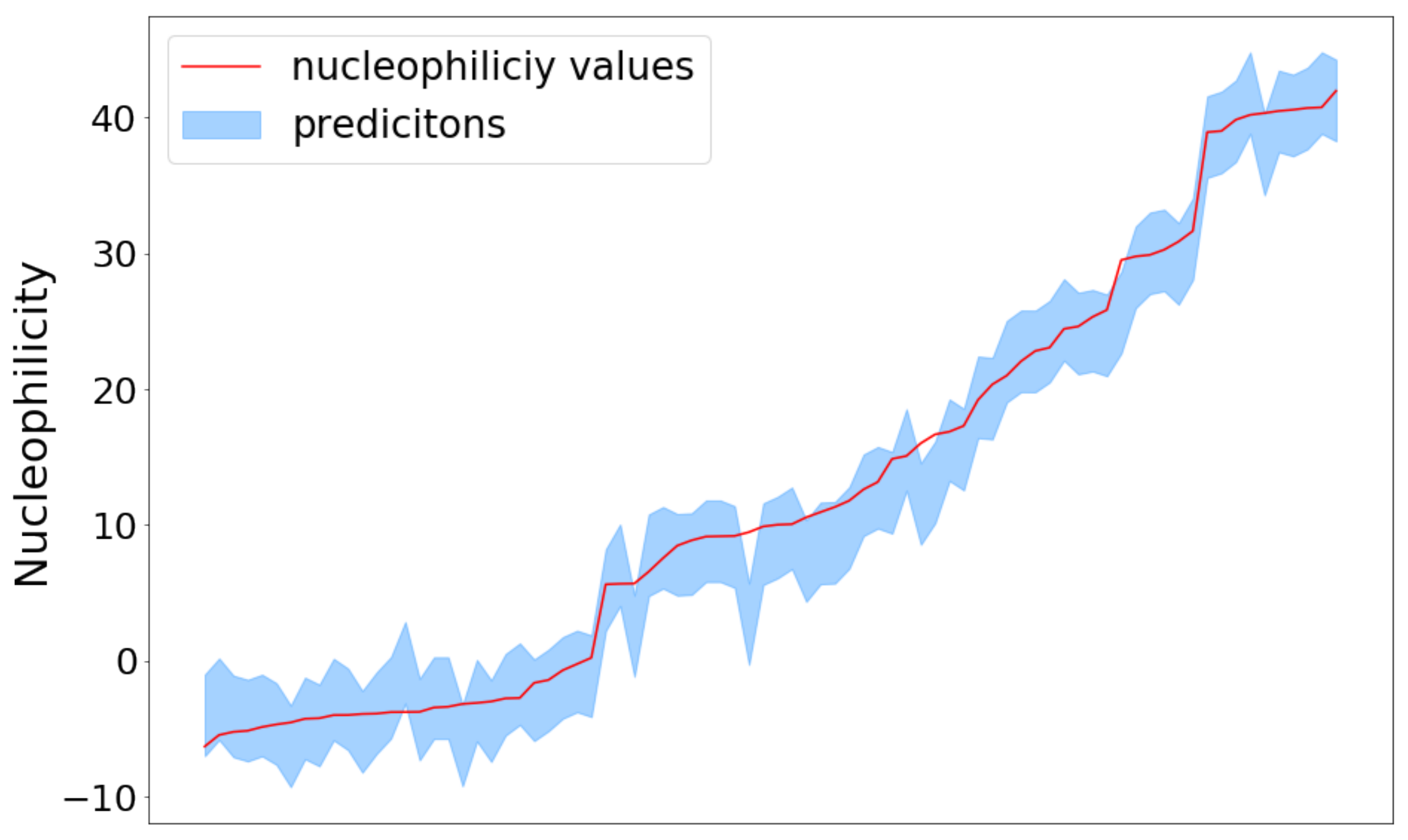}\label{fig:nuc_test}}%
    \caption{The best performance of GAT models for (a) electrophilicity and (b) nucleophilicity predictions. The red curve is the actual electrophilicity and nucleophilicity of the test samples and the blue stripe is the predicted reactivity, plus or minus three orders of magnitude.}%
    \label{fig:test_gat}%
\end{figure}

Table \ref{tab:acc_result} shows the results obtained with four different optimized neural networks with the corresponding molecular representations. Each number in Table \ref{tab:acc_result} corresponds to the average ten-fold cross-validation accuracy, together with the corresponding standard deviation, for both electrophilicity and nucleophilicity predictions. As previously mentioned, the reactivity numbers are valid within three orders of magnitude. Thus, a prediction is considered to be correct if the predicted reactivity is within three orders of magnitude of the actual MCA* or MAA*. Although the informative vector representation of atoms shows a good performance, it has several downsides and limitations. The informative features are specifically tailored for our dataset and are not generalizable to unseen molecular structures. Also, there is no guarantee that the extracted features are the best representation of atoms. Since human chemists have designed these features based on their experience, another chemist might come up with a different set of features. On the other hand, for the SMILES representation and the corresponding networks, although there are no manually tailored features, these networks do not show comparable performance in comparison to other representations. The reasons for this poor performance might be the implicit rules of writing SMILES string which convert the molecular graphs into text representations. For instance, the long dependency between two parentheses that specify a branch of atoms might not be captured with recurrent neural networks such as LSTMs. Finally, the graph representation of molecular structure does not have any of the aforementioned drawbacks and the corresponding models produce the best results. Figure \ref{fig:test_gat} illustrates the performance of the best graph attention model for both electrophilicity and nucleophilicity predictions.

%%%%%%%%%%%%%%%%%%%%%%%%%%%%%%%%%%%%%%%%%%%%%%%%%%%%%%%%%
\begin{table*}[t]
    \footnotesize
    \centering
    \caption{Accuracy results for the prediction of both electrophilicity and nucleophilicity. All accuracy figures and error bars (standard deviations) are obtained by averaging over a ten-fold cross validation experiment. The attention graph neural network outperforms the other models and representations. }
    \vspace{3mm}
    \begin{tabular}{llccc}
    \toprule
    %\hline
        &Representation&Atomic&Group&Molecular\\
    %\hline
    \midrule
        \multirow{5}{*}{\bf Electrophilicity}&Informative fingerprint& 91.04 $\pm 0.01$&-&-\\
        %\cmidrule(lr){5-7}
        &ECFP &-&-&80.21$\pm 0.51$\\
        &SMILES &-&-&73.66$\pm 0.44$\\
        &GCN &90.03$\pm 0.63$&86.97$\pm 0.72$&86.24$\pm 0.66$\\
        &GAT &\bf 91.17$\pm 0.67$&87.66$\pm 0.45$&87.93$\pm 0.79$\\
    \midrule
    %\hline
        \multirow{5}{*}{\bf Nulecophilicity}&Informative fingerprint&92.01 $\pm 0.02$ &-&-\\
        %\cmidrule(lr){2-4}
        %\cmidrule(lr){5-7}
        &ECFP &-&-&81.04 $\pm 0.41$\\
        &SMILES &-&-&72.59 $\pm 0.39$\\
        &GCN &90.91 $\pm 0.43$&86.68 $\pm 0.77$ &87.42 $\pm 0.53$\\
        &GAT &\bf 92.14 $\pm 0.52$ &87.03 $\pm 0.51$ &87.24$\pm 0.60$\\
    \bottomrule
    %\hline
    \end{tabular}
    % }
    \label{tab:acc_result}
\end{table*}
%%%%%%%%%%%%%%%%%%%%%%%%%%%%%%%%%%%%%%%%%%%%%%%%%%%%%%%%%
%%%%%%%%%%%%%%%%%%%%%%%%%%%%%%%%%%%%%%%%%%%%%%%%%%%%%%%%%

\begin{figure}[t]
    \centering
    \subfloat[][\centering]{\includegraphics[height=7.5cm, width=7.5cm]{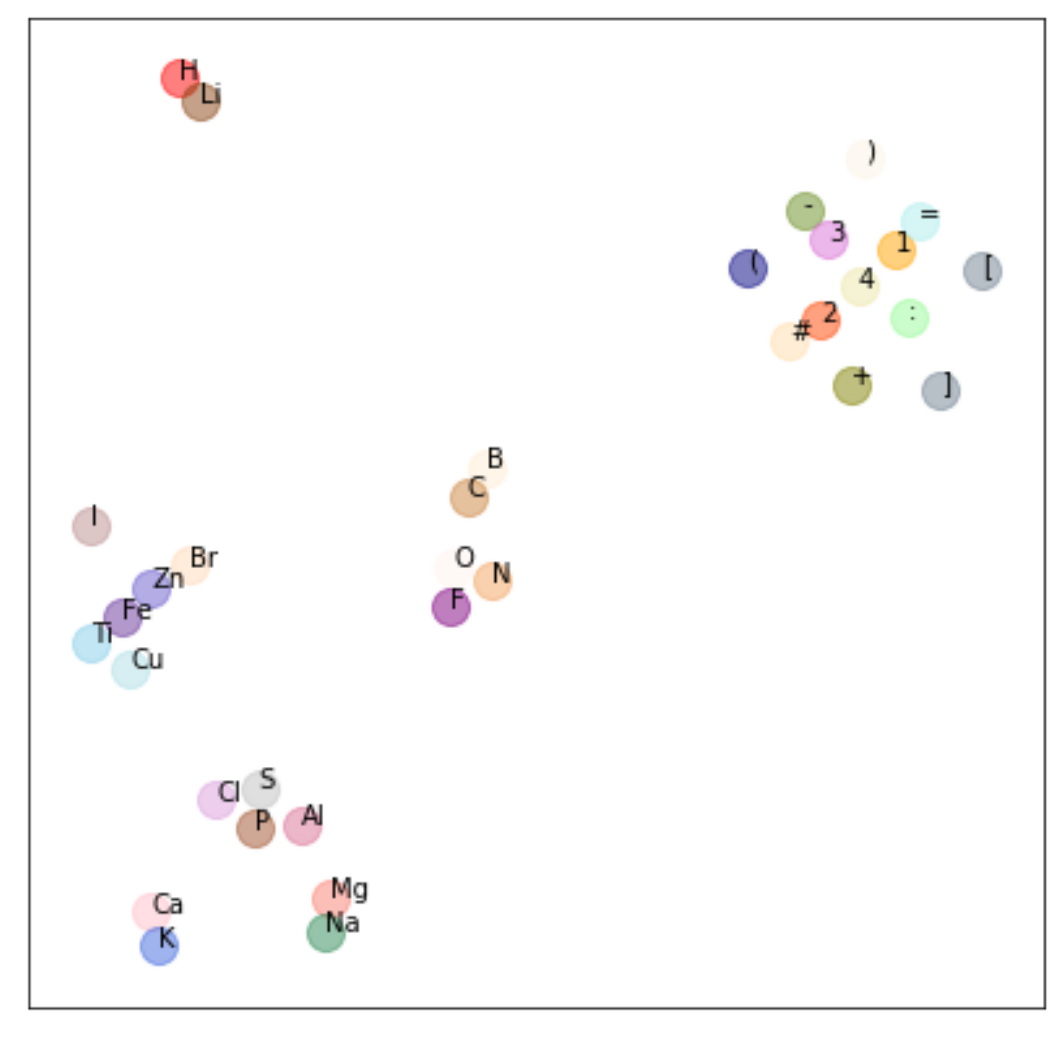}\label{fig:atom_tsne}}%
    \qquad
    \subfloat[][\centering]{\includegraphics[height=7.50cm, width=7.45cm]{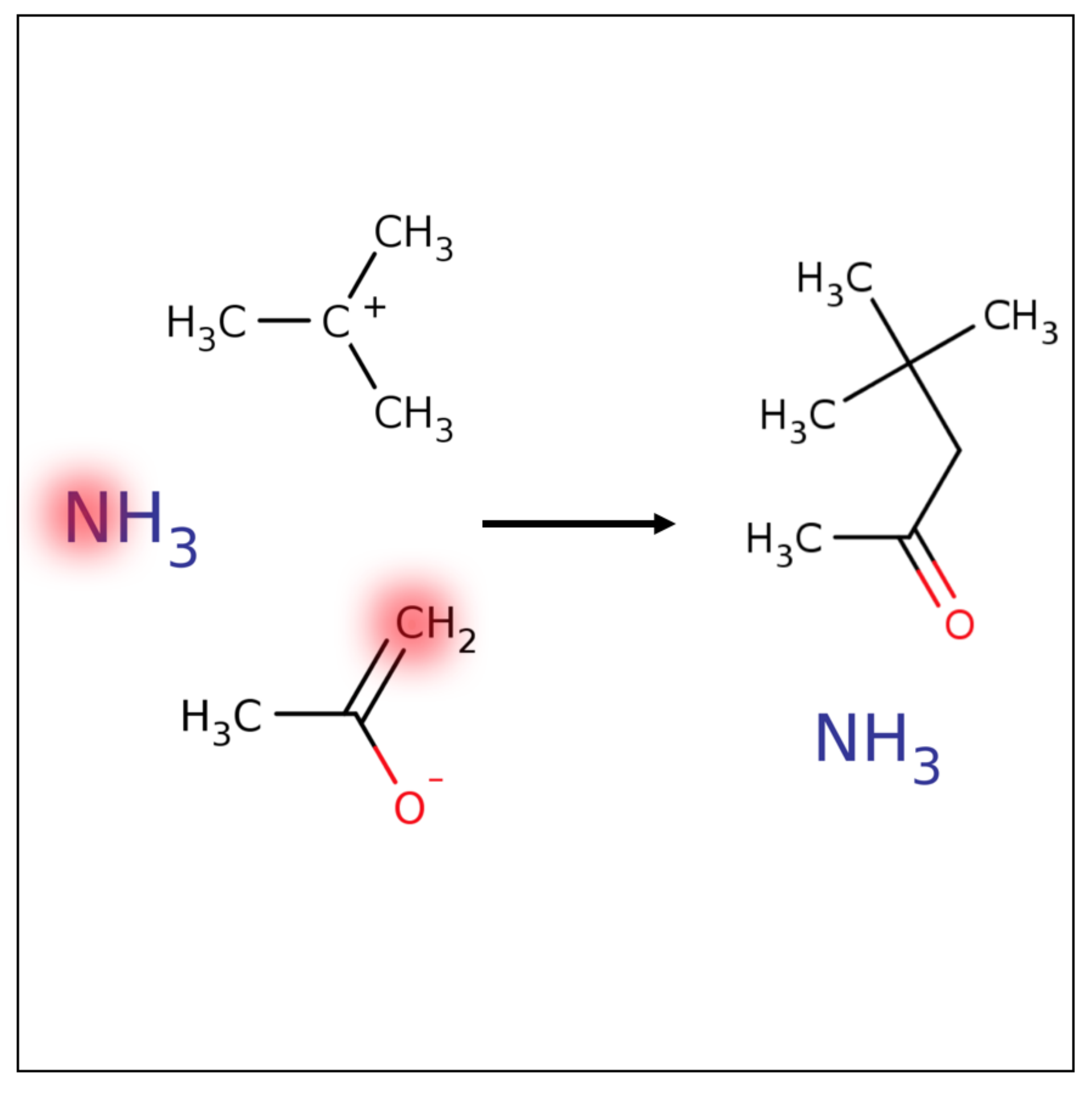}\label{fig:reaction_sample}}%
    \caption{(a): A two-dimensional t-SNE visualization of the numerical embedding for atom symbols and special characters found in SMILES strings. Special characters are clustered together and far from the atom symbols. The embeddings of atoms with similar properties are close to each other. (b): An example of a chemical reaction that can be problematic for an automated reaction prediction system. This reaction has two potential electron donors, marked in red. While both donors are reasonable, the nitrogen's reactivity score is considerably smaller than the carbon's reactivity score.}%
    \label{fig:tsne_rections_sample}%
\end{figure}

%%%%%%%%%%%%%%%%%%%%%%%%%%%%%%%%%%%%%%%%%%%%%%%%%%%%%%%%%
These results provide at least a proof-of-concept that the high-throughput prediction of the MCA* and MAA* scores is feasible. Next, we demonstrate how such a system could enable other tasks, such as chemical reaction prediction and combinatorial generation of chemical reaction mechanisms.

\vspace{2.5mm}
\subsection{Chemical Reaction Prediction}
\vspace{1.5mm}
In recent years, several deep-learning-based methods have been introduced to predict the outcome of chemical reactions under certain conditions.
These systems operate primarily either on string (SMIRKS) representations \citep{schwaller2019molecular},
or molecular graph representations \citep{kayala2011learning,kayala2012reactionpredictor,fooshee2018deep, coley2017prediction, do2019graph}. All the molecular graph-based methods seek to identify the most reactive atoms within the reactants. For instance, \citet{coley2017prediction} identify the most reactive atoms (i.e. reaction centers) using a binary classifier that classify atoms as members of the reaction center or not.  \citet{fooshee2018deep} focus on reaction mechanisms by identifying electron sources and sinks and ranking the corresponding pairings.
Although these methods yield reasonably successful reaction predictors, they are not always able to accurately identify the most reactive atoms, especially when multiple such atoms appear in one reaction. The method presented here for estimating reactivity scores for electrophiles and nucleophiles could be used to address some of these problems by using the scores to rank the atoms. Such ranking may accurately identify reaction centers and electron donors and acceptors, bypassing or complementing the complex methods described, for instance, in \citet{coley2017prediction, fooshee2018deep}. An example of this approach is depicted in Figure \ref{fig:reaction_sample}. For this example, a chemical reaction predictor such as \citet{fooshee2018deep} might predict both the nitrogen (amine group) and the highlighted carbon as the potential electron donors (since both atoms are labeled as electron donors in the training set of such system). However, our predicted nucleophilicity score for the carbon is higher than the reactivity score predicted for the nitrogen (+21.64 vs +9.71), leading to the correct identification of the electron donor and the corresponding reaction.

\iffalse
%%%%%%%%%%%%%%%%%%%%%%%%%%%%%%%%%%%%%%%%%%%%%%%%%%%%%%%%%
\begin{figure}
  \centering
  \includegraphics[width=\linewidth]{reaction_sample.pdf}
  \caption{An example of a chemical reaction that can be problematic for an automated reaction prediction system. This reaction has two potential electron donors, marked in red. While both donors are reasonable, the nitrogen's reactivity score is considerably smaller than the carbon's reactivity score.}
  \label{reaction}
\end{figure}
%%%%%%%%%%%%%%%%%%%%%%%%%%%%%%%%%%%%%%%%%%%%%%%%%%%%%%%%%
\fi

%%%%%%%%%%%%%%%%%%%%%%%%%%%%%%%%%%%%%%%%%%%%%%%%%%%%%%%%%
\begin{figure}[b!]
    \centering
    \subfloat[][\centering]{\includegraphics[width=\linewidth]{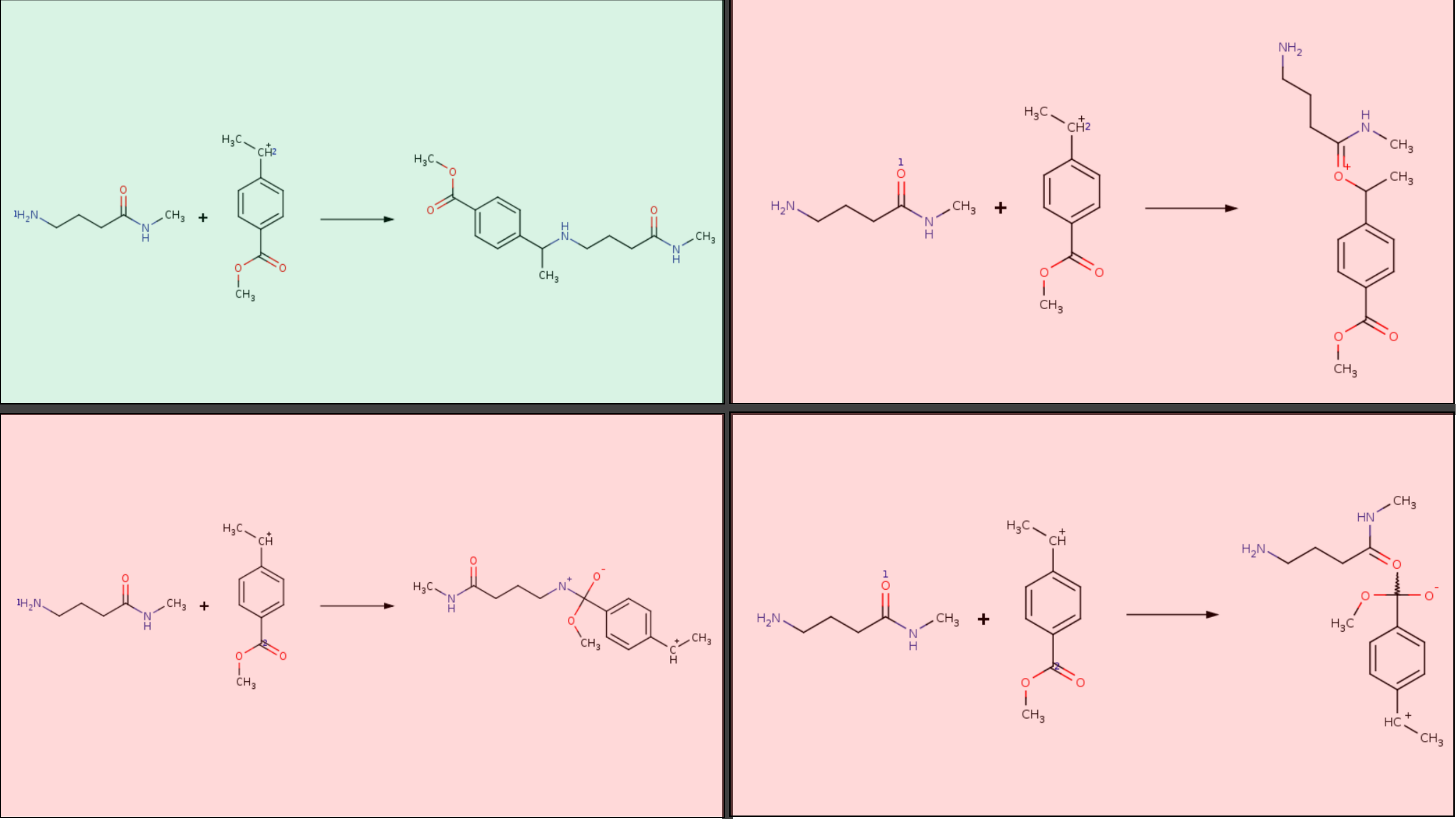}\label{fig:comb1}}%
    \qquad
    \subfloat[][\centering]{\includegraphics[width=\linewidth]{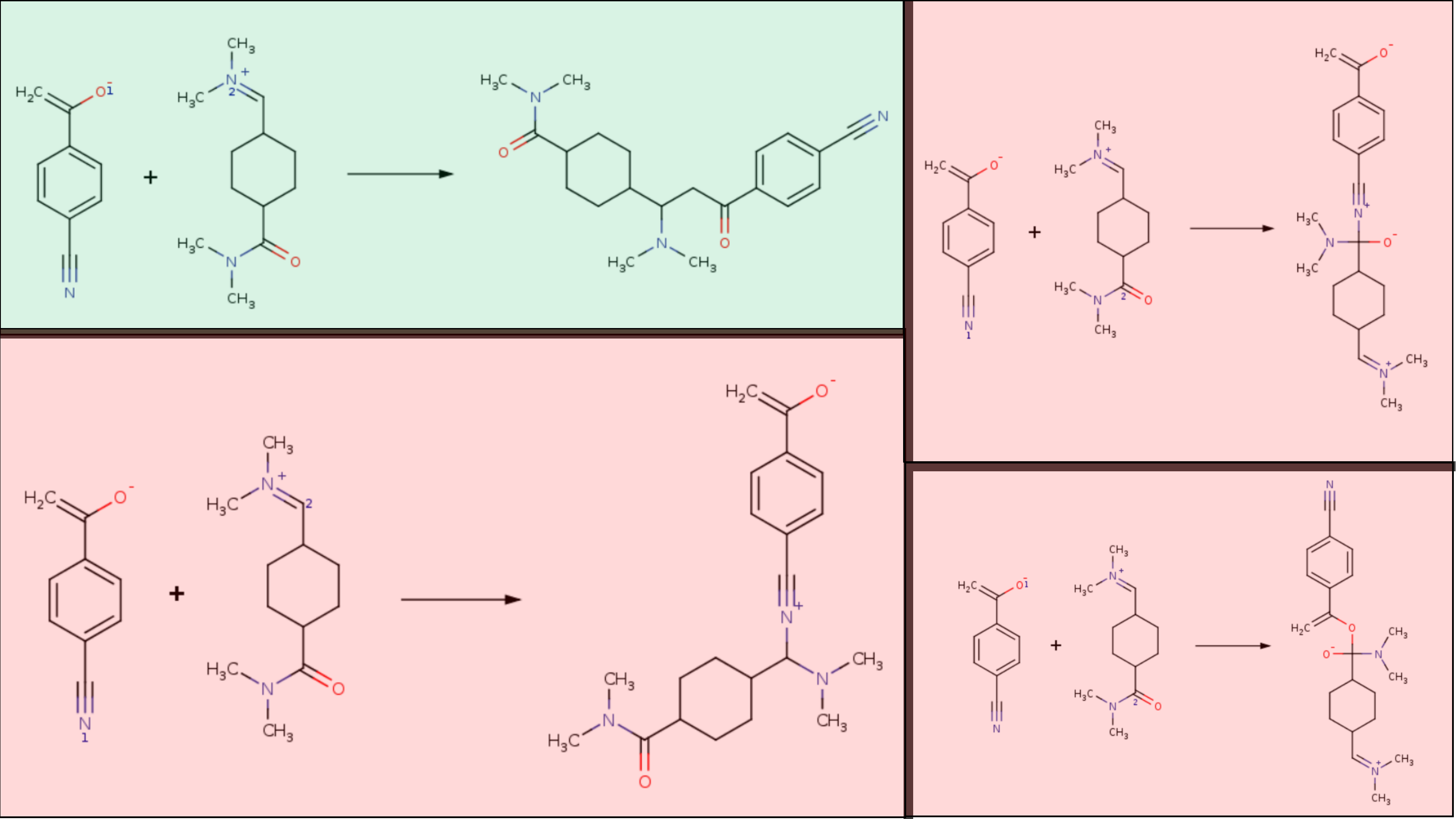}\label{fig:comb2}}%
    \caption{(a): The four possible reactions corresponding to the first row of Table \ref{tab:comb} [$T_1+S_2$; $T_2+S_1$; $T_1+T_2$; $S_1+S_2$]. (b): Likewise, the four possible reactions corresponding to the second row of Table \ref{tab:comb}. In each figure, the reaction predicted using the reactivity scores is depicted in green, whereas the remaining three less plausible reactions are depicted in red. These implausible reactions can be automatically filtered out during the process of combinatorial reaction generation.}%
    \label{fig:comb_example}%
\end{figure}

%%%%%%%%%%%%%%%%%%%%%%%%%%%%%%%%%%%%%%%%%%%%%%%%%%%%%%%%%

%%%%%%%%%%%%%%%%%%%%%%%%%%%%%%%%%%%%%%%%%%%%%%%%%%%%%%%
\begin{table*}[t!]
\footnotesize
\centering
\caption{Two examples of combinatorial reaction mechanisms. $T_1$ and $T_2$ correspond to electron donor and electron acceptor templates respectively. Similarly, $S_1$ and $S_2$ correspond to electron donor and electron acceptor substituents respectively. $E(.)$ and $N(.)$ denote the predicted electrophilicity and nucleophilicity values. The last two columns show the difference between the reactivity of templates and substituents for both electron donors and electron acceptors. The positive numbers indicate that the predicted reactivity of the templates is higher than the predicted reactivity of the substituents (see text).
}
 \vspace{3mm}
 \begin{tabular}{cccccc} 
 \toprule
 %\hline
 \multicolumn{2}{c}{Templates}&\multicolumn{2}{c}{Substituents}&&\\
 \cmidrule(lr){1-2}
 \cmidrule(lr){3-4}
 $T_1$(donor)&$T_2$(acceptor)&$S_1$(donor)&$S_2$(acceptor)& {\footnotesize$ N(T_1)-N(S_1)$}&  {\footnotesize$E(T_2)-E(S_2)$}\\
 \midrule
 %\hline
 N[R] & C[CH+][R] & O=C([R])NC & [R]C(OC)=O  & +11.47 & +28.21\\ 
 C=C([R])[O-] & C/[N+](C)=C\textbackslash[R] & N\#C[R] & O=C([R])N(C)C & +7.6 & + 18.48\\
 \bottomrule
 %\hline
 \end{tabular}
 \label{tab:comb}
\end{table*}
%%%%%%%%%%%%%%%%%%%%%%%%%%%%%%%%%%%%%%%%%%%%%%%%%%%%%%%%

\vspace{2.5mm}
\subsection{Combinatorial Generation of Chemical Reaction Mechanisms}
\vspace{2.5mm}
\citet{fooshee2018deep} introduced a method to combinatorially augment a training set of reactions. This method uses two fixed scaffolds respectively containing one electron donor group and one electron acceptor group (also called templates) and then combinatorially varies the decorative atoms (also called substituents) attached to the templates within realistic chemical constraints. Following this process, one can generate large numbers of elementary reactions covering a range of fundamental reaction classes.  The most important constraint to enable this method to generate plausible reactions is that the atoms within the attaching substituent functional groups must be less reactive than the template atoms. Automatically enforcing this constraint requires a ranking of atoms based on their reactivity within different functional groups, and this can be done using the scales proposed here. A demonstration is given in Table \ref{tab:comb} and Figure \ref{fig:comb_example} showing that the proposed scales can be used to filter out the non-plausible combinatorially generated mechanisms. 
The first row of Table \ref{tab:comb} and Figure \ref{fig:comb1} consider four functional groups in two reactants. The higher nucleophilicity of the amino group over the carboxamide and the higher electrophilicity of the carbocation over the ester carbonyl correctly predict the attack of the amino group on the carbocation (the green reaction in Figure \ref{fig:comb1}). Similarly, the second row of Table \ref{tab:comb} and Figure \ref{fig:comb2} consider four other functional groups in two more reactants. The higher nucleophilicity of the enolate relative to the nitrile and the higher electrophilicity of the iminium ion over the carboxamide carbonyl group correctly predicts a Mannich reaction involving the addition of the enolate to the iminium ion (the green reaction in Figure \ref{fig:comb2}).

\section{Conclusion}

Methyl cation affinity and methyl anion affinity have been shown to be highly correlated to the reactivity of atoms in functional groups over a broad range of organic chemistry. Leveraging this correlation, we used DFT calculations to curate a dataset of relative reactivity scores for 2,421 electrophilic and nucleophilic functional groups covering 53 orders of magnitude of chemical reactivity.
This curated dataset is available to the community and was used to train several deep neural networks, with different representations, to estimate reactivity. Through experiments, we have shown that graph attention neural networks outperform other methods and representations and can accurately estimate the reactivity with a ten-fold cross-validation accuracy of 92\% showcasing another synergistic application of QM and Machine Learning methods

In the future, it may be useful to incorporate the experimentally validated reactivity parameters from the Mayr database and to further expand the curated dataset of DFT-derived reactivity scores (MCA* and MAA*) to a broader range of molecular structures. All the proposed methods in this work are trained using the 2421 structures with decent coverage of basic functional groups over 53 orders of magnitude of chemical reactivity. However, the method in \citet{mca2021, mood2020methyl} for calculating chemical reactivity is applicable to molecular structures with a chemical reactivity covering 180 orders of magnitude. Thus, there is significant room for curating larger data sets and training more refined machine learning systems to predict chemical reactivity, as well as using the machine learning results to guide the DFT calculations towards the most informative regions of chemical reaction space.  

\section{Acknowledgement}
Work in part supported by NSF grants
1633631 and 195811 to DVV and PB.

\bibliography{main}

\end{document}